\documentclass[10pt,prd,superscriptaddress,preprintnumbers,amsmath,amssymb,nofootinbib]{revtex4}
\usepackage{epsfig}  
\usepackage{graphicx}
\usepackage{hyperref}
\usepackage{color}
\usepackage{float}
\usepackage{amsfonts}
\usepackage{amsmath}
\usepackage{slashed}
\usepackage{soul}

\large

\begin{document} 

\title{Flavor signatures of complex anomalous  $tcZ$ couplings}

\author{Suman Kumbhakar}
\email{suman@phy.iitb.ac.in}
\affiliation{Indian Institute of Technology Bombay, Mumbai 400076, India}

\author{Jyoti Saini}
\email{saini.1@iitj.ac.in}
\affiliation{Indian Institute of Technology Jodhpur, Jodhpur 342037, India}

\begin{abstract}

In this work we study the effects of anomalous $tcZ$ couplings. Such couplings would potentially affect several neutral current decays of $K$ and $B$ mesons via $Z$-penguin diagrams. Using constraints from relevant observables in $K$ and $B$ sectors, we find that the 2$\sigma$ upper bound on the branching ratio of ${\cal B}(t \to c Z)$ is $1.47 \times 10^{-5}$ for real coupling and $1.91 \times 10^{-4}$ for complex coupling. The current experimental upper bound from ATLAS and CMS collaborations on the branching ratio of $t \to c Z$ are $2.4 \times 10^{-4}$ and $4.9\times 10^{-4}$ at $95\%$ C.L., respectively. Hence the possibility of observation of $t \to c Z$ decay at the level of $10^{-4}$  would imply the anomalous couplings to be complex. Such  complex  couplings should also show up its presence in other related decays. We find that an order of magnitude enhancement is possible in the branching ratio of $K_L \to \pi^0 \nu \bar{\nu}$. Further, the complex $tcZ$ coupling can also provide large enhancements in many $CP$ violating angular observables in $B \to K^* \mu^+ \mu^-$ decay.


\end{abstract}

\maketitle 

\section{Introduction}
The currently running experiments in the LHC have reported a plethora of measurements in $B$ meson decays. Some of these measurements related to the decays induced by the quark level transition $b \to s\, \mu^+ \,\mu^-$ do not agree with the predictions of the  Standard Model (SM). These deviations are not statistically significant to claim discovery of new physics. However they can be considered as hints of beyond SM physics. These measurements are
: \footnote{Apart from introducing new anomalies in $b \to s\, \mu^+ \,\mu^-$ sector, the LHC reinforced the prevailing anomalies in the decays included by $b \to c \tau \nu$ transition. A series of experiments performed by  BaBar \cite{Lees:2012xj,Lees:2013uzd}, Belle \cite{Huschle:2015rga,Sato:2016svk,Hirose:2016wfn,Hirose:2017dxl} indicated an excess in the values of the ratios $R_{D^{(*)}} = {\Gamma(B\rightarrow D^{(*)}\,\tau\,\bar{\nu})}/{ \Gamma(B\rightarrow D^{(*)}\, e/\mu \, \bar{\nu})}$ over their SM predictions. This was corroborated by the LHCb measurements \cite{Aaij:2015yra,Aaij:2017uff}.  Refs.  \cite{Freytsis:2015qca,Alok:2017qsi,Alok:2018uft,Blanke:2018yud,Alok:2019uqc,Murgui:2019czp} identified new physics operators which can account for $R_{D^{(*)}}$ measurements. }
\begin{itemize}
\item  In 2014, the LHCb collaboration reported the measurement of the ratio $R_K \equiv  \Gamma(B^+ \to K^+ \,\mu^+\,\mu^-)/\Gamma(B^+ \to K^+\,e^+\,e^-)$ to be $0.745^{+0.090}_{-0.074} \, (\rm stat.)\pm 0.036 \, (\rm syst.)$ in the di-lepton invariant mass-squared $q^2$ range ($1.0 \le q^2 \le 6.0 \, {\rm GeV}^2$) \cite{rk} which deviates from the SM prediction of $\approx 1$ \cite{Hiller:2003js,Bordone:2016gaq,Bouchard:2013mia}. To be specific, the measurement deviates from the SM value of $ 1.00\pm 0.01$ \cite{Bouchard:2013mia} by 2.6$\sigma$. Recently in Moriond'19, LHCb has updated the measurement of $R_K$ from the Run-II data. The updated value of $R_K$ is $0.846^{+0.060}_{-0.054}\,(\rm stat.)^{+0.016}_{-0.014}\,{\rm syst.}$~ \cite{Humair} which is  still $\simeq 2.5\sigma$ away from the SM. This measurement  is considered to be an indication of lepton flavor universality violation in the $b \to s l^+ l^-$ sector.
  
\item The measurement of $R_K$  was further corroborated by the measurement of $R_{K^*} \equiv \Gamma (B^0 \to K^{*0} \mu^+\mu^-)/\Gamma(B^0 \to K^{*0} e^+ e^-)$. This ratio was measured  in the low ($0.045 \le q^2 \le 1.1 \, {\rm GeV}^2$) as well as in the central ($1.1 \le q^2 \le 6.0 \, {\rm GeV}^2$) $q^2$ bins. The measured values are $0.660^{+0.110}_{-0.070}\,(\rm stat.)\pm 0.024\,(\rm syst.)$ for low $q^2$ and $0.685^{+0.113}_{-0.069}\,(\rm stat.)\pm 0.047\,(\rm syst.)$ for central $q^2$~\cite{rkstar} which deviates from the SM predictions \cite{  Hiller:2003js,Bordone:2016gaq,Bouchard:2013mia,Jager:2014rwa,Straub:2015ica,Serra:2016ivr,Capdevila:2017ert}. 
The measured values of $R_{K^*}$ differ from the SM predictions of $R_{K^*}^{\rm low} = 0.906\pm 0.028$ and $R_{K^*}^{\rm central} = 1.00\pm 0.01$ \cite{Bordone:2016gaq}  by $\sim 2.5\sigma$ and $\sim 3\sigma$, in the low and central $q^2$ regions, respectively.
In Moriond'19, the Belle collaboration has presented their first measurements of $R_{K^*}$ in both $B^0$ and $B^+$ decays~\cite{belle-rks,Abdesselam:2019wac}. These measurements are reported
in multiple $q^2$ bins and have comparatively large uncertainties. Hence the discrepancy in $R_{K^*}$ still stands at $\sim 2.4\sigma$ level. The  measurements of $R_{K^{(*)}}$ could be an indication of presence of NP in $b \to s \mu^+ \mu^-$ and/or $b \to s e^+ e^-$ sector.

\item The experimentally measured values of some of the angular observables in $B \to K^* \mu^+ \mu^-$ \cite{Kstarlhcb1,Kstarlhcb2,KstarBelle} do not agree with their SM predictions \cite{sm-angular}\footnote{This can also be attributed to underestimation of hadronic uncertainties within the SM. For e.g., in \cite{Ciuchini:2015qxb}, it was shown that the experimental data can be accommodated within the SM itself if one assumes sizable non factorizable power corrections.} In particular, the angular observable $P'_5$ disagrees with the SM at the level of 4$\sigma$ in the  $q^2$ bin $4.3-8.68$ GeV$^2$.  This disagreement is confirmed by the recent ATLAS \cite{Aaboud:2018krd} measurements.  However, the CMS \cite{Sirunyan:2017dhj} collaboration has measured $P'_5$ in multiple $q^2$ bin and the measured values are consistent with the SM predictions. Further, the  measured value of the branching ratio of $B_s \to \phi \mu^+ \mu^-$ \cite{bsphilhc1,bsphilhc2} disagrees with the SM prediction at the level of of 3$\sigma$. The discrepancies in $P'_5$ and the branching ratio of $B_s \to \phi \mu^+ \mu^-$ can be attributed to the presence of NP in  $b \to s \, \mu^+ \, \mu^-$ sector only.  Hence it would be natural to account for  all of these anomalies by assuming NP only in $b \to s \mu^+ \mu^-$ transition.  In recent times, many groups have identified the Lorentz structure of new physics  which can account for all measurenets in the  $b \to s \, \mu^+ \, \mu^-$ sector \cite{rkstar-before,rkstar-after}.
\end{itemize}

Apart from the decays of $B$ meson, the top quark decays are particularly important for hunting physics beyond the SM. As it is the heaviest of all the SM particles, it is expected to feel the effect of NP most. Also, LHC is primarily a top factory producing abundant top quark events.  Hence one expects the observation of possible anomalous couplings in the top sector at the LHC.
The flavor changing neutral current (FCNC) top quark decays, such as $t\to q Z$ $(q=u,c)$, are highly suppressed within the SM  owing to the GIM mechanism. The SM predictions for the branching ratios of $t \to  uZ$ and $t \to cZ$ decays are $\sim$ $10^{-17}$ and $10^{-14}$, respectively \cite{Eilam:1990zc,AguilarSaavedra:2004wm}, and are beyond the scope of current detection level at the LHC  until NP enhances their branching ratios up to the present sensitivity of LHC. The present upper bound on the branching ratio of $t \to c Z$ are $2.4 \times 10^{-4}$ \cite{Aaboud:2018nyl} from ATLAS and $4.9\times 10^{-4}$~\cite{Sirunyan:2017kkr} from CMS collaboration at $95\%$ C.L. Thus the observation of any of the FCNC top quark decays at the LHC would imply discovery of NP.

The possible NP effects in the top FCNC decays have been incorporated in a model independent way within different theoretical frameworks, see for e.g., \cite{Hewett:1993em,Han:1995pk,Han:1996ep,Beneke:2000hk,Liu:2005dp,misiak,Ferreira:2008cj,Zhang:2008yn,Coimbra:2008qp,Li:2011af,Gong:2013sh,Durieux:2014xla,Nardi:1995fq,Chala:2018agk,Banerjee:2018fsx,Khanpour:2014xla}. In this work we investigate the flavour signatures  of effective anomalous $t \to cZ$  couplings. Apart from enhancing  the branching ratio of $t \to c Z$ decay, such couplings would also affect the loop level processes involving the top quark, and hence has the potential to affect rare $B$ and $K$ meson decays via the $Z$ penguin diagrams.
 The anomalous $tcZ$ couplings can arise in various extensions of the SM. These include  universal extra-dimensional model \cite{Chiang:2018oyd}, Randall-Sundrum framework of warped extra dimension \cite{Agashe:2006wa}, type-III two-Higgs doublet model \cite{Gaitan:2017tka}, and models with vector-like quarks \cite{AguilarSaavedra:2002kr}.

In this work we use measurements of  (i) all CP conserving observables in $b\to s \mu^+ \mu-$ sector, (ii) observables in $b\rightarrow s e^+ e^-$ sector, (iii) the branching ratio of $B^+ \to \pi^+ \mu^+ \mu^-$ and $B_d \to \mu^+ \mu^-$ and (iv) the branching fraction of $K^+ \to \pi^+\, \nu \, \bar{\nu}$  to perform a combined fit to the anomalous $tcZ$ coupling. In doing the analysis, we consider the coupling to be real as well as complex. For the complex $tcZ$ coupling,  we find that the $2\sigma$ upper bound of the branching ratio of $t \to c\, Z$ is $1.91\times 10^{-4}$.
Then we look for other flavor signatures of this anomalous $tcZ$ coupling. In particular, we examine branching ratio of $K_L\to \pi \, \nu \, \bar{\nu} $ and various $CP$ violating angular observables in ${B} \to (K, {K}^*)\, \mu^+\, \mu^-$. We find that the complex $tcZ$ coupling can give rise to large enhancements to various $CP$ violating observables.
 
This paper is organized as follows. In sec. II, we describe the effect of $tcZ$ coupling on rare $B$ and $K$ decays. In sec. III, we discuss about the methodology used in the fit. In sec. IV, we provide fit results along with predictions of various observables. We present our conclusions in sec. V.

\section{Effect of anomalous $t\to cZ$ couplings on rare $B$ and $K$ decays}
\label{tcz}

The effective $tcZ$ Lagrangian can be written as~\footnote{In Ref.~\cite{Durieux:2014xla},  all possible FCNC transitions of  top quark was considered. It was shown that there could be a large number of gauge invariant terms which can lead to different top quark FCNC processes. In this work, we concentrate only on the effective tcZ anomalous couplings and obtain model independent bounds using constraints from the current flavor data. In the presence of other anomalous top quark FCNC couplings, the allowed tcZ parameter space, in general may be relaxed.  Further, in a specific new physics model there can be additional constraints which may shrink the allowed tcZ parameter space obtained in this work.} \cite{AguilarSaavedra:2008zc}
\begin{equation}
{\cal L}_{tcZ} = \frac{g}{2 \cos \theta_W} \bar{c}\gamma^{\mu} \left(g^L_{ct} P_L + g^R_{ct} P_R\right)tZ_{\mu}  +\frac{g}{2 \cos \theta_W} \bar{c}\frac{i\sigma^{\mu \nu}p_{\nu}}{M_Z} \left(\kappa^L_{ct} P_L
+ \kappa^R_{ct} P_R\right)tZ_{\mu} + {\rm h.c.},
\label{Leff}
\end{equation}
where $P_{L,R}\equiv(1\mp\gamma_5)/2$ and $g^{L,R}_{ct}$ and $\kappa^{L,R}_{ct}$ are NP couplings.

The rare $B$ meson decays induced by the quark level transitions, such as $b \to s\, l^+\, l^-$ and $b \to d\, l^+\, l^-$
along with rare $K$ meson decays induced by  the quark level transition, such as $s \to d \nu \bar{\nu}$, 
occur only at the loop level within the SM and hence are highly suppressed. 
Within the SM, these processes are dominated  by box and $Z$ penguin diagrams.
The anomalous $tcZ$ couplings can enter into the $Z$ penguin diagrams and hence has the potential to affect these decays. 
Therefore these decays can be used to constrain anomalous $t\to cZ$ coupling.

Let us now consider the contribution of anomalous $tcZ$ couplings to 
the rare $B$ decays induced by the quark level transition $b \to s\,  \mu^+\, \mu^-$. 
Within the SM, the effective Hamiltonian for the quark-level transition $b \to s\, \mu^+\, \mu^- $ can be written as 
\begin{eqnarray}
\mathcal{H}_{SM} &=& − \frac{4 G_F}{\sqrt{2} \pi} V_{ts}^* V_{tb} \left[ \sum_{i=1}^{6} C_i(\mu) \mathcal{O}_i(\mu) + C_7 \frac{e}{16 \pi^2} \left[\overline{s} \sigma_{\mu \nu}\left(m_s P_L \right.\right.\right. \nonumber\\ 
& & \left. \left. +m_b P_R\right)b\right]F^{\mu \nu} + C_9 \frac{\alpha_{em}}{4 \pi}\left(\overline{s} \gamma^{\mu} P_L b\right)\left(\overline{\mu} \gamma_{\mu} \mu\right) + C_{10} \frac{\alpha_{em}}{4 \pi} \nonumber\\  
& & \left. \left(\overline{s} \gamma^{\mu} P_L b\right)\left(\overline{\mu} \gamma_{\mu} \gamma_5 \mu\right) \right],
\label{Heffbsmumu}
\end{eqnarray}
where $G_F$ is the Fermi constant, $V_{ts}$ and $V_{tb}$ are the Cabibbo-Kobayashi-Maskawa (CKM) matrix elements and $P_{L,R} = (1 \mp \gamma^{5})/2$ are the projection operators. The effect of the operators $O_i,\,i=1-6,8 $ can be embedded in the redefined effective Wilson coefficients (WCs) as $C_7(\mu)\rightarrow C^{\rm eff}_7(\mu,q^2)$ and $C_9(\mu)\rightarrow C^{\rm eff}_9(\mu,q^2)$, where the scale $\mu = m_b$.  The form of the operators $\mathcal{O}_i$  are given in ref.~\cite{Buchalla:1995vs}. The effective $tcZ$ vertices, given in eq.~(\ref{Leff}), modifies the WCs $C_{9}$ and $C_{10}$
 through the $Z$ penguin diagrams. There are two such diagrams, one of which is proportional to $V_{tb}V^{*}_{cs}$ whereas the other is proportional to $V_{cb}V^{*}_{ts}$. As  $V_{tb}V^{*}_{cs}$ $\sim$ $O(1)$ and  $V_{cb}V^{*}_{ts}$ $\sim$ $O(\lambda^4)$, the dominant contribution comes from the diagram proportional to $V_{tb}V^{*}_{cs}$ and  the contribution from the other diagram can be safely neglected. The NP contributions to $C_{9}$ and $C_{10}$  are given as \cite{Li:2011af,Gong:2013sh}

\begin{equation}
{C}_9^{s, NP} = - {C}_{10}^{s, NP} =   - \frac{1}{8 \sin^2 \theta_W} \frac{V^*_{cs}}{V^*_{ts}} \Bigg[\left(-x_t \ln \frac{M^2_W}{\mu^2}+\frac{3}{2}+x_t-x_t\ln x_t\right)g^L_{ct}\Bigg],
\label{npbs}
\end{equation}
with $x_t =\bar{m}^2_t/M^2_W$. Here the right handed coupling, $g^R_{ct}$, is neglected as it is suppressed by a factor of $\bar{m_c}/M_W$. The NP contributions to $C_{9,\,10}$ have been calculated in the unitary gauge with the modified minimal subtraction ($\rm \overline{MS}$) scheme \cite{Li:2011af}. It is found that the NP tensors operators, defined in the effective Lagrangian Eq.~(\ref{Leff}), do not contribute to $C_{9,\,10}$.

The effective Hamiltonian for the process $b \to d\, \mu^+\, \mu^-$ can be obtained from eq.~(\ref{Heffbsmumu}) by replacing $s$ by $d$. The contribution of anomalous $tcZ$ coupling to $b \to d\,  \mu^+ \,\mu^-$ transition modifies the WCs $C_{9}$ and $C_{10}$. The new contributions to these WCs are

\begin{equation}
{C}_9^{d, NP} = - {C}_{10}^{d, NP} =- \frac{1}{8 \sin^2 \theta_W} \frac{V^*_{cd}}{V^*_{td}} \Bigg[\left(-x_t \ln \frac{M^2_W}{\mu^2}+\frac{3}{2}+x_t-x_t\ln x_t\right)g^L_{ct}\Bigg].
\label{cnpb2d}
\end{equation}
Here we have neglected the contributions from right handed coupling, $g^R_{ct}$, which is suppressed by a factor of $\bar{m}_c/M_W$. Further, the contribution due to the CKM suppressed Feynman diagram is also neglected.

We now consider NP contribution to $s \to d \, \nu \bar{\nu}$ transition.  
 In this case, the
CKM contribution from both the diagrams is of the same order, $O(\lambda^3)$,
and hence we include both of them in our analysis. The
$K^+ \to \pi^+ \, \nu \bar{\nu}$ decay is the only observed decay channel 
induced by the quark level transition $\bar{s} \to \bar{d} \, \nu \bar{\nu}$.
Unlike other $K$ decays, the SM prediction for the branching ratio of $K^+ \to \pi^+ \, \nu \bar{\nu}$ is under good control as the long distance contribution to ${\cal B}(K^+\to \pi^+\nu\bar{\nu})$ is about three orders of magnitude smaller than the short-distance 
contribution \cite{Rein:1989tr,Hagelin:1989wt}.
The effective Hamiltonian for $K^+ \to \pi^+ \, \nu \bar{\nu}$ in the SM can be written as
\begin{widetext}
\begin{equation}
{\cal H}_{eff} = \frac{G_F}{\sqrt{2}}\frac{\alpha}{2 \pi \sin^2 \theta_W} \sum_{l=e,\mu,\tau} 
\left[V^*_{cs}V_{cd} X^l_{NL} + V^*_{ts}V_{td} X(x_t) \right]
\times(\bar{s}d)_{V-A}(\bar{\nu}_l\nu_l)_{V-A}\,
\end{equation}
\end{widetext}
where $X^l_{NL}$ and $X(x_t)$ are the structure functions corresponding to charm and top sector, respectively 
\cite{Buchalla:1993wq,Buchalla:1995vs,Buchalla:1998ba}. 
The contribution of anomalous $tcZ$ coupling to 
$\bar{s} \to \bar{d} \, \nu \bar{\nu}$ transition then 
 modifies the structure function $X(x_t)$ in the following way 
\begin{equation}
X(x_t) \rightarrow X^{\rm tot}(x_t) = X(x_t) + X^{NP},
\label{xtot}
\end{equation}
where
\begin{eqnarray}
X(x_t)&=&\eta_X \frac{x_t}{8}\left[\frac{2+x_t}{x_t-1}+\frac{3x_t-6}{(1-x_t)^2}\ln x_t\right],\\
X^{NP} &=& -\frac{1}{8} \left(\frac{V_{cd}V^*_{ts}+V_{td}V^*_{cs}}{V_{td}V^*_{ts}}\right) \left(-x_t \ln \frac{M^2_W}{\mu^2}+\frac{3}{2}+x_t-x_t\ln x_t\right)(g^{L}_{ct})^*\,.
\end{eqnarray}
Here $\eta_x=0.994$ is the NLO QCD correction factor.

\begin{table}[htbp]
\begin{center}
\begin{tabular}{|c|c|}
\hline
$G_F = 1.16637 \times 10^{-5}$ Gev$^{-2}$& $\tau_{B_d}= (1.519\pm 0.007)$ ps   \\
$ \sin^2\theta_w = 0.23116$              & $\tau_{B^+}=1.638\pm 0.004$ ps  \\
$ \alpha(M_Z) = \frac{1}{129}$         &  $f_{B_d}=(190 \pm 1.3 )$ MeV \cite{Aoki:2019cca}   \\
$ \alpha_s(M_Z) = 0.1184$                &  $ \kappa_+ = (5.36\pm0.026)\times 10^{-11}$ \cite{Mescia:2007kn} \\
$m_t(m_t) = 163 $ GeV                    & $ \kappa_L=(2.31\pm 0.01)\times 10^{-10}$ \\
 $M_W = 80.385$ GeV                      & $\lambda=0.225\pm 0.001$ \cite{UTfit} \\
        $M_Z = 91.1876 $ GeV                     & $A=0.826\pm 0.012$ \cite{UTfit} \\
      $M_{B_d} = 5.27917$ GeV            & $\bar{\rho}= 0.148\pm 0.013$ \cite{UTfit}\\
 $M_{B^+} = 5.27932$ GeV                      & $\bar{\eta}= 0.348\pm 0.010$ \cite{UTfit} \\
$m_{\mu} = 0.105$ GeV                     & \\
\hline
\end{tabular}
\caption{Decay constants, bag parameters, QCD corrections and other
  parameters used in our analysis.  When not explicitly stated, they are taken from the Particle Data Group \cite{pdg}.}
\label{tab-input}
\end{center} 
\end{table}

\section{Constraints on the anomalous $tcZ$ couplings}
\label{constraint}

In order to obtain the constraints on the anomalous $tcZ$ coupling $g^L_{ct}$, 
we perform a $\chi^2$ fit using all measured observables in $B$ and $K$ sectors. The $\chi^2$ fit is performed using the CERN minimization code {\tt MINUIT} \cite{James:1975dr}. The total $\chi^2$ is written as a function of two parameters: ${\rm Re} (g^L_{ct})$, 
 and ${\rm Im} (g^L_{ct})$.
The $\chi^2$ function is defined as 
\begin{equation}
\chi^2_{\rm total} =
\chi^2_{b \to s\, \mu^+ \,\mu^-} 
+\chi^2_{b \to s\, e^+ \,e^-} 
+\chi^2_{b \to d\, \mu^+ \,\mu^-}  
+ \chi^2_{s \to d \nu \bar{\nu}}~.
\end{equation}

In our analysis, $\chi^2$ of an observable $A$ is defined as
\begin{equation}
\chi^2_A = \left( \frac{A - A_{exp}^c}{A_{exp}^{err}} \right)^2,
\end{equation}
where the measured value of $A$ is $(A_{exp}^c \pm A_{exp}^{err})$.

In the following subsections, we discuss the individual components of the function $\chi^2_{\rm total}$, 
i.e the $\chi^2$ of different observables which are being used as inputs. 

\subsection{Constraints from $b \to s\, \mu^+ \,\mu^-$ sector}
As the anomalous $tcZ$ coupling considered in this work does not account for lepton flavour non-universal effects, we do not consider $R_K$ and $R_{K^*}$ measurements in our fits. Instead, we consider all relevant measurements related  to $b \to s\, \mu^+ \,\mu^-$ and $b \to s\, e^+ \,e^-$.

The quark level transition $b \to s\, \mu^+\, \mu^-$ induces inclusive and exclusive semi-leptonic $B$ decays along with the purely leptonic ${\bar B}_s \to \mu^+ \mu^-$ decay. 
In our analysis we include following observables:
\begin{enumerate}

	\item The branching ratio of $B_{s} \to \mu^{+}\mu^{-}$ which is $(3.1 \pm0.7) \times 10^{-9}$ \cite{Amhis:2016xyh}.
        
        \item The differential branching ratio of $B^0 \to K^{*0} \mu^+ \mu^-$ measured in various $q^2$ bins \cite{Aaij:2016flj,CDFupdate,Chatrchyan:2013cda,Khachatryan:2015isa}.  
        
        \item Various angular observables in $B^0 \to K^{*0} \mu^+ \mu^-$ \cite{Aaboud:2018krd,Sirunyan:2017dhj,Khachatryan:2015isa,Kstarlhcb2,CDFupdate}. 
        
        \item The differential branching ratio of $B^{+} \to K^{*+}\mu^{+}\mu^{-}$ in various  $q^2$ bins \cite{Aaij:2014pli,CDFupdate}.

	\item The differential branching ratio of $B^{0}\rightarrow K^{0}\mu^{+}\mu^{-}$ in various $q^2$ bins \cite{Aaij:2014pli,CDFupdate}. 
	
	\item The differential branching ratio of $B^{+}\rightarrow K^{+}\mu^{+}\mu^{-}$ in several $q^2$ bins \cite{Aaij:2014pli,CDFupdate}.
	
		\item  The measurements of the angular observables and the differential branching ratio  of $B^{0}_{s}\to \phi \mu^{+}\mu^{-}$ \cite{bsphilhc2}.

       \item The experimental measurements for the differential branching ratio of $B \to X_{s}\mu^{+}\mu^{-}$ \cite{Lees:2013nxa}.

 \end{enumerate}
 
In the context of anomalous $tcZ$  couplings, the WCs are related as ${C}_9^{s, NP} = - {C}_{10}^{s, NP}$. Hence we can use the above $b \to s \mu^+ \mu^-$ observables to obtain constraints on the coupling $g^L_{ct}$ using eq.~(\ref{npbs}).

The  $\chi^2$  function for all $b \to s \mu^+ \mu^-$ observables listed above  is defined as
\begin{equation}
\chi^2_{b \to s\, \mu^+ \,\mu^-} = (O_{th}(C_i) -O_{exp})^T \, \mathcal{C}^{-1} \,
(O_{th}(C_i) -O_{exp})\,
\end{equation} 
where $O_{th}(C_i)$ are the  theoretical predictions of $b \to s \mu^+ \mu^-$ observables which are calculated using {\tt flavio} \cite{Straub:2018kue} and $O_{exp}$ are the corresponding experimental measurements.   
The total covariance matrix $\mathcal{C}$ is obtained by adding the individual theoretical and experimental covariance matrices.

\subsection{Constraints from $b \to s\, e^+ \,e^-$ sector}
 The following measurements, mediated by $b \to s\, e^+ \,e^-$ transitions, are taken into fit:
\begin{enumerate}
\item Angular observables $P'_4$ and $P'_5$ in $B\rightarrow K^*\,e^+\,e^-$, both in two $q^2$ bins $[1,6]$ and $[14.18,19]$ GeV$^2$~\cite{Wehle:2016yoi}.
\item The branching ratio of $B^0\rightarrow K^*\,e^+\,e^-$ in $[0.001,1]$ GeV$^2$ bin~\cite{Aaij:2013hha}.
\item The branching ratio of $B^+\rightarrow K^+\,e^+\,e^-$ in [1,6] GeV$^2$ bin~\cite{Aaij:2019wad}.
\item The $K^*$ polrization fraction in $B^0\rightarrow K^*\,e^+\,e^-$~\cite{Aaij:2015dea}.
\item  The branching fractions of $B\rightarrow X_s\,e^+\,e^-$ within two $q^2$ bins $[1,6]$ and $[14.2,25]$ GeV$^2$~\cite{Lees:2013nxa}.
\end{enumerate} 
 
The constraints coming from $B_s\rightarrow e^+\,e^-$  is weak as the current upper bound on its branching ratio \cite{Amhis:2016xyh} is  about seven  orders of magnitude away from the SM prediction. Although this is not included directly  in our fit, we have checked that the fit results do not evade the upper limit. The $\chi^2$ function for the set of observables in $b \to s\, e^+ \,e^-$ transition is defined as
\begin{equation}
\chi^2_{b \to s\, e^+ \,e^-} = \sum \frac{ \left(O_i^{th} - O^{exp}_i\right)^2}{\sigma^{th\,2}_{O_i}+\sigma^{exp\,2}_{O_i}},
\end{equation} 
where $\sigma^{th}_{O_i}$ and $\sigma^{exp}_{O_i}$ are the SM and experimental uncertainties in each observable.
\subsection{Constraints from $b \to d\, \mu^+ \,\mu^-$ sector}

The quark level transition $b \to d \mu^+\, \mu^-$ gives rise to inclusive semi-leptonic decay ${\bar B} \to X_d \,\mu^+ \, \mu^-$ and exclusive semi-leptonic decay such as ${\bar B} \to (\pi^0,\,\rho)\,\mu^+ \,\mu^-$. However, so far, none of these decays have been observerd. We only have an upper bound on their branching ratios \cite{Wei:2008nv,Lees:2013lvs}. However, LHCb has observed the $B^+ \to \pi^+\, \mu^+\, \mu^-$ decay, which is induced by $b \to d \mu^+\, \mu^-$ transition, with measured branching ratio of $ (2.3 \pm 0.6 \pm 0.1) \times 10^{-8}$ \cite{LHCb:2012de}. This is the first measurement of any decay channel in  $b \to d \,\mu^+\, \mu^-$ sector. 

The theoretical expression for ${\cal B}(B^+ \to \pi^+\, \mu^+\, \mu^-)$ 
in the presence of anomalous $tcZ$ coupling can be obtained from Ref.~\cite{Wang:2007sp} by replacing the WCs $C_{9,\,10}$ by that given in eq.~(\ref{cnpb2d}). The contribution to $\chi^2_{\rm total}$ is
\begin{equation}
\chi^2_{B^+ \to \pi^+\, \mu^+\, \mu^-}  =\Big( \frac{{\cal B}(B^+ \to \pi^+\, \mu^+\, \mu^-) - 2.3\times 10^{-8}}
{0.66\times 10^{-8}} \Big)^2\; ,
\end{equation}
where, following ref.~\cite{Wang:2007sp}, a theoretical error of $15\%$ is included in ${\cal B}(B^+ \to \pi^+\, \mu^+\, \mu^-)$. This error is mainly due to uncertainties in the $B^+ \to \pi^+$ form factors \cite{Ball:2004ye}.

We also include the constraint from $B_d \to \mu^+ \mu^-$ decay. The branching ratio of $B_d \to \mu^+ \,\mu^-$ in the presence of anomalous $tcZ$ coupling is
given by
\begin{eqnarray}
\mathcal{B}(B_d \to \mu^+ \,\mu^-) &=& \frac{G^2_F \alpha^2 M_{B_d} m_\mu^2 f_{B_d}^2 \tau_{B_d}}{16 \pi^3} 
|V_{td}V^*_{tb}|^2 \times \nonumber\\
& & \sqrt{1 - 4 (m_\mu^2/M_{B_d}^2)} 
\left|{C}_{10} + C^{d,NP}_{10} \right|^2 .
\end{eqnarray}
In order to include ${\cal B}(B_d \to \mu^+ \,\mu^-)$  in the
fit, we define
\begin{equation}
B_{\rm lepd} = \frac{16 \pi^3 {\cal{B}}( B_d \to \mu^+ \,\mu^-)}{G^2_F \alpha^2 M_{B_d} m_\mu^2 f_{B_d}^2 \tau_{B_d}
 |V_{td}V^*_{tb}|^2 \sqrt{1 - 4 (m_\mu^2/M_{B_d}^2)}} \,.
\end{equation}
Using the inputs given in Table~\ref{tab-input} and ${\cal B}(B_d \to \mu^+ \,\mu^-)_{\rm exp}= (3.9\pm 1.6)\times 10^{-10}$~\cite{Amhis:2016xyh} , we get
\begin{equation}
B_{\rm lepd, {\rm exp}} = 60.83 \pm 25.18.
\end{equation}
The contribution to $\chi^2$ from  ${\cal B}(B_d \to \mu^+ \,\mu^-)$ is then
\begin{equation}
\chi^2_{B_d \to \mu^+ \mu^-} = \Big( \frac{B_{\rm lepd} - 60.83}{25.18} \Big)^2\;.
\end{equation}
Thus we have,
\begin{equation}
\chi^2_{b \to d\, \mu^+\, \mu^-} = \chi^2_{B^+ \to \pi^+\, \mu^+\, \mu^-} + \chi^2_{B_d \to \mu^+ \mu^-}\,.
\end{equation}

\subsection{Constraints from $s \to d\,  \nu {\bar \nu}$ sector}
The branching ratio of $K^+ \to \pi^+ \nu {\bar \nu}$, the only measurement in this sector, in the presence of anomalous $tcZ$ coupling is given by
\begin{widetext}
\begin{equation}
\frac{{\cal B}(K^+ \to \pi^+ \nu {\bar \nu})}{\kappa_{+}} =
\left( \frac{{\rm Re}(V_{cd}V^*_{cs})}{\lambda} P_c(X) +\frac{{\rm Re}(V_{td}V^*_{ts})}{\lambda^5}X^{\rm tot}(x_t) \right)^2 +
\left(\frac{{\rm Im}(V_{td}V^*_{ts})}{\lambda^5} X^{\rm tot}(x_t)\right)^2 ,
\end{equation}
\end{widetext}
where $P_c(X)=0.38\pm 0.04$ \cite{Buras:2006gb} is the NNLO QCD corrected structure 
function in the charm sector and
\begin{equation}
\kappa_+ = r_{K^+} \frac{3 \alpha^2 {\cal B}(K^+\to \pi^0 e^+ \nu)}{2 \pi^2 \sin^4 \theta_W} \lambda^8 ~.
\end{equation}
Here $r_{K^+}=0.901$ encapsulates the isospin-breaking corrections in
relating the branching ratio of $K^+\to \pi^+\nu\bar{\nu}$ to that of
the well-measured decay $K^+\to \pi^0 e^+ \nu$. $X^{\rm tot}(x_t)$ is given in
eq.~(\ref{xtot}).
With inputs used in table~\ref{tab-input} and ${\cal B}(K^+\to \pi^+\nu\bar{\nu})_{\rm exp}$, we estimate
\begin{equation}
\frac{{\cal B}(K^+\to \pi^+\nu\bar{\nu})}{\kappa_+} = 3.17 \pm 2.05 ~,
\end{equation} 

In order to include
${\cal B}(K^+\to \pi^+\nu\bar{\nu})$ in the fit, we define
\begin{eqnarray}
\chi^2_{K^+\to \pi^+\nu\bar{\nu}} &=&\Big( \frac{{\cal B}(K^+\to \pi^+\nu\bar{\nu})/\kappa_{+} - 3.17}{2.05} \Big)^2 \nonumber\\
& &+\Big( \frac{P_c(X) - 0.38}{0.04} \Big)^2 .
\end{eqnarray}
Thus, the error on $P_c(X)$ has been taken into account by considering it
to be a parameter and adding a contribution to $\chi^2_{\rm total}$.
\section{Results and discussions}
\label{res}
The fit results for real and complex $tcZ$ couplings are presented in Table \ref{fit-res}. 

\begin{table*}[htbp]
\begin{center} 
	\tabcolsep 3pt
 \begin{tabular}{|c | c |} 
 \hline
Real coupling & Complex coupling  \\ 
 \hline\hline
$g^L_{ct}=(-6.51 \pm 1.46) \times 10^{-3}$ & ${\rm Re} (g^L_{ct}) = (-1.02 \pm 0.38) \times 10^{-2}; \,
{\rm Im} (g^L_{ct}) = (1.79 \pm 0.84) \times 10^{-2}$ \\ 
 \hline
\end{tabular}
\caption{\label{fit-res} Values of anomalous $tcZ$ couplings.}
\end{center}
\end{table*}
\subsection{${\cal B}(t\to c\, Z)$  }
The branching ratio of $t \to c Z$ is defined as~\cite{Han:1995pk,Beneke:2000hk,Bernreuther:2008ju}
\begin{equation}
{\cal B}(t\to c\, Z)= \frac{\Gamma(t\rightarrow c Z)}{\Gamma(t\rightarrow b W)},
\end{equation}
where, the leading-oder (LO) decay width of $t\rightarrow b W$ is given as~\cite{Li:1990qf}
\begin{equation}
\Gamma (t\rightarrow b W) = \frac{G_F m^3_t}{8\sqrt{2}\pi} \vert V_{tb}\vert^2 \beta^4_W (3-2\beta^2_W),
\end{equation}
with $\beta_W = (1-m^2_W/m^2_t)^{1/2}$, being the velocity of the $W$ boson in the top quark rest frame. On the other hand, the LO decay width of $t\rightarrow c Z$ can be written as~\cite{Han:1995pk,Beneke:2000hk,Bernreuther:2008ju}
\begin{equation}
\Gamma (t\rightarrow c Z) = \frac{G_F m^3_t}{8\sqrt{2}\pi}\frac{\vert g_{ct}^L\vert^2 + \vert g_{ct}^R\vert^2}{2}\beta^4_Z (3-2\beta^2_Z),
\end{equation}
where $\beta_Z = (1-m^2_Z/m^2_t)^{1/2}$, the velocity of the $Z$ boson in the top quark rest frame.

Using the fit results, we find that for real $tcZ$ coupling, ${\cal B}(t \to c \, Z)= (7.73 \pm 3.47) \times 10^{-6}$ and the $2\sigma$ upper bound is $1.47\times 10^{-5}$. For complex $tcZ$ coupling, the branching ratio of $t\rightarrow c\,Z$ is $(7.75\pm 5.67)\times 10^{-5}$ and the 2$\sigma$ upper bound on the branching ratio is $1.91 \times 10^{-4}$ which could be measured at the LHC. In fact this is very close to the present upper bound from ATLAS~\cite{Aaboud:2018nyl} and CMS~\cite{Sirunyan:2017kkr} collaborations. The ATLAS collaboration reported the upper limit to be $2.4\times 10^{-4}$ at $95\%$ C.L. using the data collected at a center of mass energy of $13$ TeV. The CMS collaboration set the upper limit at $4.9\times 10^{-4}$ at $95\%$ C.L. using the data recorded at a center of mass energy of 8 TeV. Thus the observation of $t \to c Z$ at the level of  $10^{-4}$ would not only be an indication of anomalous $tcZ$ coupling but could also imply that these couplings are complex. 
If such $tcZ$ complex couplings are present then it should also show its presence in other decays. In the following we study the impact of this coupling on various $CP$-violating observables related to the rare decays of $K$ and $B$ mesons.  It would be interesting to see whether large deviations are possible in some of these  observables.

\subsection{${\cal B}(K_L\to \pi^0 \, \nu \, \bar{\nu})$  }

The branching ratio of $K_L\to \pi^0 \, \nu \, \bar{\nu}$  is a purely $CP$ violating quantity, i.e., it vanishes if $CP$ is conserved. The SM branching ratio is predicted to be $(2.90 \pm 0.40)\times 10^{-11}$.  As this process is highly suppressed in the SM, it is sensitive to new physics. The preset upper bound set by KOTO experiment at J-PARC on ${\cal B}(K_L\to \pi^0 \, \nu \, \bar{\nu})$ is $3.0\times 10^{-9}$~\cite{Shiomi:2018qip} at $90\%$ C.L. which is about three orders of magnitude above the SM prediction. This result improves the previous upper limit~\cite{Ahn:2009gb,Shiomi:2014sfa} by an order of magnitude. This experiment should have enough data for the first observation of the decay by about 2021.

Within the SM, $K_L\to \pi^0 \, \nu \, \bar{\nu}$ decay is dominated by loop diagrams with top quark exchange. Hence anomalous $tcZ$ coupling can modify its branching ratio. The branching ratio of $K_L \to \pi^0 \, \nu \, \bar{\nu}$ in the presence of $tcZ$ coupling is given by 
\begin{equation}
{\cal B}(K_L\to \pi^0 \, \nu \, \bar{\nu}) = \kappa_L \left[\frac{{\rm Im}\left(V_{ts}^* V_{td} X^{\rm tot}(x_t)\right)}{\lambda^5}\right]^2,
\end{equation}
where $X^{\rm tot}(x_t)$ is given in Eq.~(\ref{xtot}). 

Using fit result for the complex $tcZ$ coupling, we get $\mathcal{B}(K_L \to \pi^0 \nu \bar{\nu})=(8.83 \pm 4.68) \times 10^{-11}$. The  2$\sigma$ upper bound on ${\cal B}(K_L\to \pi^0 \, \nu \, \bar{\nu})$ is obtained to be  $\leq 1.82\times 10^{-10}$. Thus the anomalous $tcZ$ couplings can provide an order of magnitude enhancement in the branching  ratio of  $K_L\to \pi^0 \, \nu \, \bar{\nu}$.
\begin{figure*}[htbp]
\centering
\resizebox{\textwidth}{!}{ 
\begin{tabular}{ccc}
\includegraphics[]{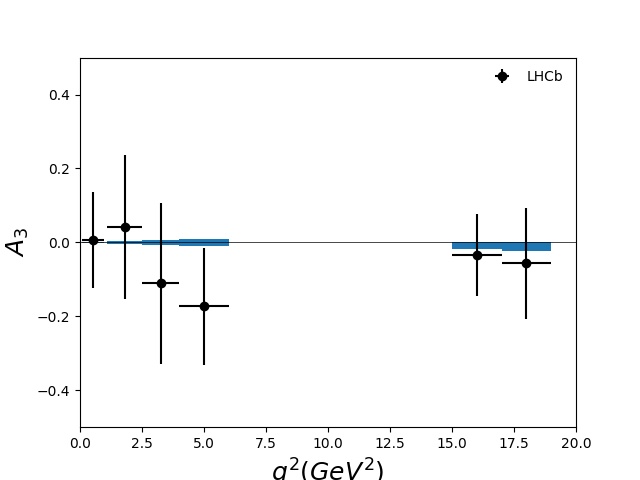} & \includegraphics[]{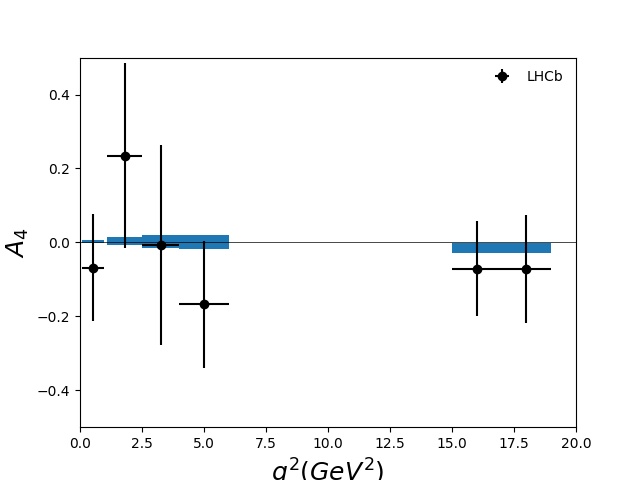} &\includegraphics[]{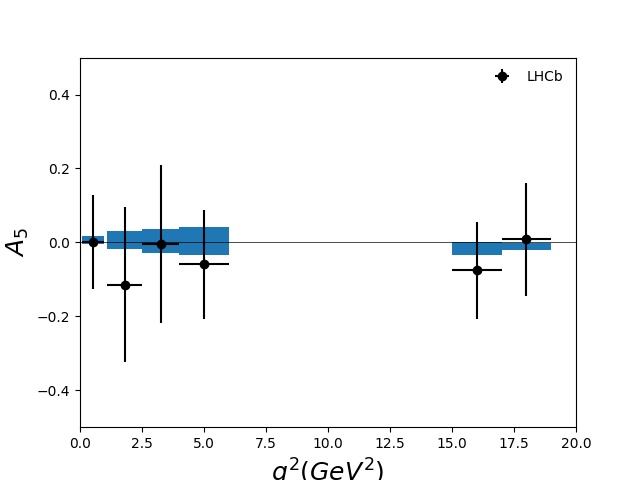} \\
\includegraphics[]{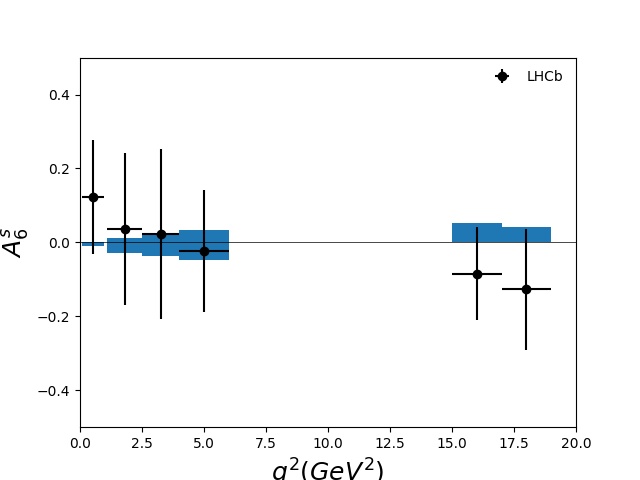} &\includegraphics[]{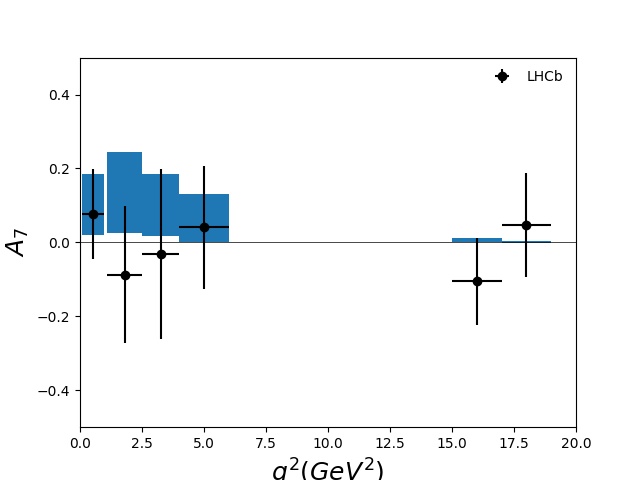} & \includegraphics[]{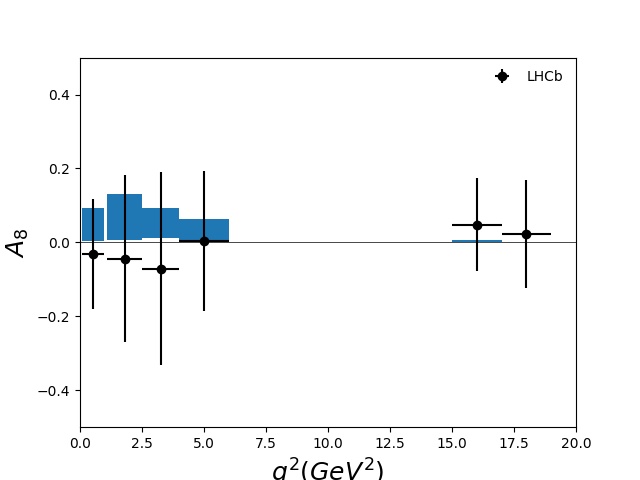}\\
\includegraphics[]{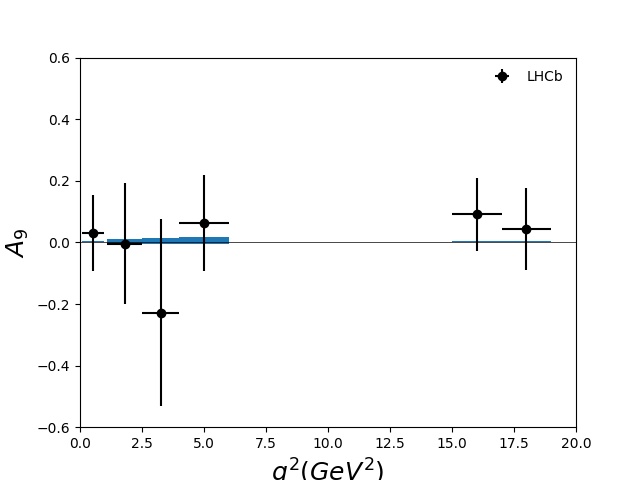}&\includegraphics[]{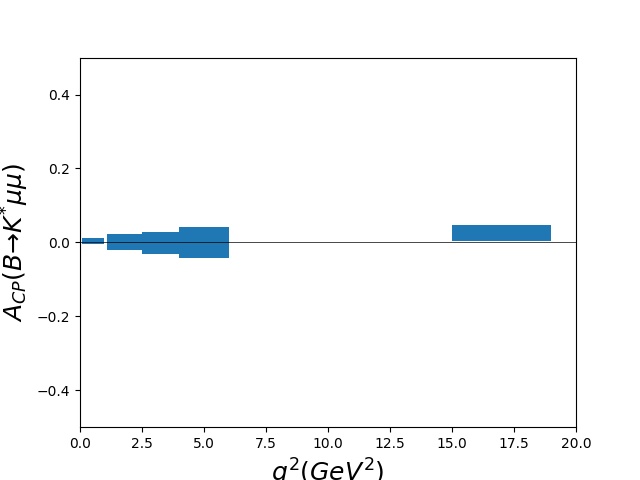}& \includegraphics[]{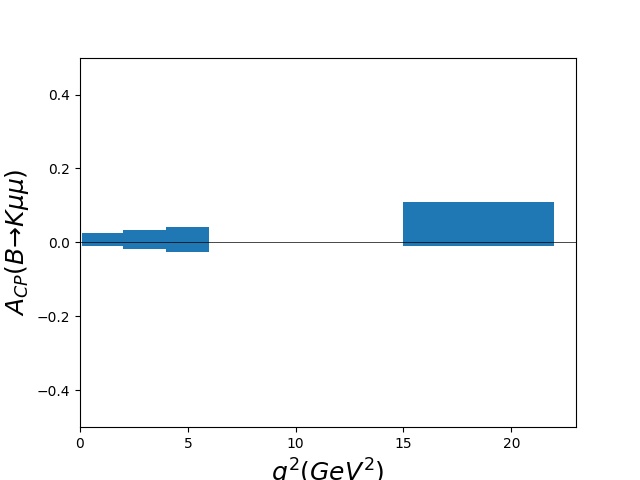}
\end{tabular}
}
\caption{(Color Online) The  plots depicts various $CP$ violating observables in  $B \to (K^*,\,K)\, \mu^+ \,\mu^-$ decays as a function of $q^2$.}
\label{fig-cpv1}
\end{figure*}

\subsection{$CP$ violating  observables in $B \to (K^*,\,K)\, \mu^+ \,\mu^-$}
 Within the SM, the $CP$ violating observables in $b \to s\, \mu^+ \,\mu^-$ are highly suppressed. 
 The complex $tcZ$ coupling can give rise to additional phase for $CP$ violation, which can affect  several $CP$ violating observables in $b \to s\, \mu^+ \,\mu^-$ sector. Here, we study various $CP$ violating observables in $B \to (K^*,\,K)\, \mu^+ \,\mu^-$ decays in the presence of complex anomalous  $tcZ$ couplings.  The four fold angular distribution of the decay $\bar{B^0} \to \bar{K}^{*0}(K^-\pi^+)\mu\mu$ can be described by the kinematical variables $q^2, \theta_{K},\theta_l$ and $\phi$~\cite{Altmannshofer:2008dz}
\begin{equation}
\frac{d^4 \Gamma(\bar{B^0} \to \bar{K}^{*0}(K\pi)\mu\mu)}{dq^2\,d\cos{\theta_l}\,d\cos{\theta_K}\,d\phi} = \frac{9}{32\pi}I(q^2,\theta_l, \theta_K, \phi),
\label{fulldist}
\end{equation}
where
\begin{widetext}
\begin{eqnarray}
I(q^2,\theta_l,\theta_K, \phi) &=& I_1^s\sin^2{\theta_K}+I_1^c\cos^2{\theta_K}+(I_2s\sin^2{\theta_K}+I_2^c\cos^2{\theta_K})\cos{2\theta_l}+I_3\sin^2{\theta_K}\sin^2{\theta_l}\cos{2\phi}\nonumber\\
&&+I_4\sin{2\theta_K}\sin{2\theta_l}\cos{\phi}+I_5\sin{2\theta_K}\sin{\theta_l}\cos{\phi}+(I_6^s\sin^2{\theta_K}+I_6^c\cos^2{\theta_K})\cos{\theta_l}\nonumber\\
&&+I_7\sin{2\theta_K}\sin{\theta_l}\sin{\phi}+I_8\sin{2\theta_K}\sin{2\theta_l}\sin{\phi}+I_9\sin^2{\theta_K}\sin^2{\theta_l}\sin{2\phi}.
\end{eqnarray}
\end{widetext}
The corresponding full angular distribution of the CP-conjugate decay $B^0 \to K^{*0}(K^+\pi^-)\mu\mu$ can be obtained by replacing the angular coefficients $I_{1,2,3,4,7} \to \bar{I}_{1,2,3,4,7}$ and $I_{5,6,8,9}$ $\rightarrow -\bar{I}_{5,6,8,9}$ in Eq.~(\ref{fulldist}) and $\bar{I_i}$ is equal to $I_i$ with all weak phase conjugated. 

The CP-violating observables are defined as
\begin{equation}
A_i = \frac{I_i - \bar{I_i}}{d(\Gamma + \bar{\Gamma})/dq^2}
\end{equation}
These asymmetries are largely suppressed in SM because of the small weak phase of CKM and hence they are  sensitive to complex NP couplings.
These symmetries can get significant contribution from the NP in the presence of CP-violating phase  \cite{Alok:2008dj,Alok:2011gv,Altmannshofer:2013foa}. The $CP$ asymmetries for $B\rightarrow (K,K^*)\,\mu^+\,\mu^-$ decays have been measured at the level of few percent at the LHCb \cite{LHCb:2012kz,Aaij:2013dgw,Aaij:2014bsa}. Measurement of other $CP$ violating observables would require higher statistics which can be achieved at HL-LHC~\cite{Cerri:2018ypt}.

Along with the LHCb measurements, the predictions for $CP$ violating asymmetries $A_i$ in the presence of complex anomalous $tcZ$ couplings are shown in fig.~\ref{fig-cpv1}. 
 In ref. \cite{Bobeth:2008ij}, it was shown that the
asymmetries $A_3$ and $A_9$ are very sensitive to the chirally flipped operators. In the present scenario, there is no chirally flipped operator contribution and hence  large enhancement is not expected in these 
asymmetries. This can also be seen from our results.  The 
asymmetry $A_7$ can be enhanced upto $~20\%$ whereas enhancement in $A_8$ can be upto $~10\%$ in the low-$q^2$ region. We also find that 
large entrancement is not possible the direct-$CP$ asymmetries in $B \to K\mu^+\mu^-$ and $B \to K^* \mu^+ \mu^-$.

\section{Conclusions}
\label{concl}
 The FCNC decays of top quark are considered as a reliable probe to physics beyond SM. Although these transitions are highly suppressed in the SM, but the promising new physics contributions can enhance them to the observation level of current collider experiments. In this work, we study the flavour signatures of effective anomalous $t\rightarrow c Z$ couplings. We use all relevant measurements in $B$ and $K$ meson systems to constrain the new physics parameter space by considering both real and complex $tcZ$ couplings. Finally, we provide predictions for the branching ratio of $t\rightarrow c Z$ along with branching ratio of $K_L \rightarrow \pi^0\nu\bar{\nu}$ and various $CP$ violating observables in $B\rightarrow (K, K^*) \mu^+\mu^-$ decays. Our finding are as follows:
\begin{itemize}
\item For real $tcZ$ coupling, the $2\sigma$ upper bound on the branching ratio of $t\rightarrow c Z$ is $1.47\times 10^{-5}$, whereas that for the complex coupling is $1.91\times 10^{-4}$. The current experimental upper bound on the branching ratio of $t\rightarrow c Z$ is $2.4\times 10^{-4}$. Hence, any future measurement of this branching ratio at the level of $10^{-4}$ would imply the coupling to be complex.

\item The complex $tcZ$ coupling provide an interesting signature in the case of  $CP$ violating decay $K_L \rightarrow \pi^0\nu\bar{\nu}$. We find that the $2\sigma$ upper bound on the branching ratio of this decay is $1.82\times 10^{-10}$, an order of magnitude higher than its SM prediction.

\item The complex $tcZ$ couplings can provide large enhancements in various $CP$ violating observables in $B \to K^*\, \mu^+ \,\mu^-$ decay.   

\end{itemize}

\section*{Acknowledgments}
We thank A. K. Alok, Dinesh Kumar, Jacky Kumar and Gauhar Abbas for  useful discussions. We are thankful to the referee for his/her comments and suggestions regarding the manuscript. 

\end{document}